\documentclass[final,5p,times]{elsarticle}
\usepackage{amsmath}
\usepackage{amssymb}
\usepackage{graphicx}
\newcommand{\dfr}[2]{\frac {\displaystyle #1}{\displaystyle #2}}

\begin{document}

\begin{frontmatter}

\title{Theoretical description of slow non-monotonic relaxation processes in Al--Y melts}

\author[l1,l2]{Vasin\,M.\,G.}
\ead{dr\underline{ }vasin@mail.ru}
\author[l1,l2]{Menshikova S.\,G.}
\author[l1]{Ivshin M.\,D.}

\address[l1]{Physical-Technical Institute of UrB RAS, 426000 Izhevsk, Russia}
\address[l2]{Institute for High Pressure Physics of RAS, 142190, Moscow, Russia}

%\ead{dr_vasin@mail.ru}
\begin{abstract}
The slow non-monotonic relaxation processes, which have been recently fixed in Al--Y melts, are described theoretically. The theoretical description is based on the Cahn--Hilliard theory and functional methods of non-equilibrium dynamics. In terms of the suggested approach the reasons of this relaxation kinetics are non-linearity of the system near to the liquidus line, which sharply increases with Y concentration, and strong initial heterogeneity of the melt on the concentration of Y atoms.
According to our analysis one can conclude that the non-monotonic temporal dependence of viscosity is caused by the Ostwald ripening processes in the rich in yttrium areas.
\end{abstract}


\end{frontmatter}

\section{Introduction}
\label{intro}

Some well known physical phenomena, observed in metallurgical processes, raise questions to physicists so far.
The reason of these questions is absence of an reliable description of these phenomena in terms of a generally accepted theory. One of these phenomena is the slow non-monotonic relaxation processes in glass-forming metal melts after melting \cite{Zam,Lad}. In the aluminium melts with small Y or Ni impurity the relaxation time reaches few hours. In metallurgy these effects are explained as the result of slow dissolution of refractory solid phase fragments in liquid.
However, the kinetics of these relaxation processes cannot be explained in terms of the linear diffusion model, since the characteristic diffusive relaxation times should be of the order of some seconds. In order to verify this we can carry out the estimation of the time of diffusive dissolution of the initial inhomogeneity with linear size $L$:  $\tau^{*} \sim {L^2}/{D}$, where $D$ is the diffusion coefficient. For Y in liquid Al   $D\sim 7\cdot 10^{-9}$ m$^2$/s, and the characteristic size of  Al$_3$Y inclusions in the initial Al solid phase is $L\sim 10^{-5}$ m.
Therefore $\tau^{*}\sim 10^{-10}\cdot 10^{9}/7\sim 10^{-2}$ s. While in the experiment the relaxation time of these processes is $\sim 10^{4}$ s. (in the considered case $t_{rel}\approx 155$ min. (Fig.\,\ref{A0})), that are several orders of magnitude greater than the characteristic time of diffusion \cite{Zam,Lad}.

\begin{figure}[h!]
   \centering
   \includegraphics[width=7.1cm]{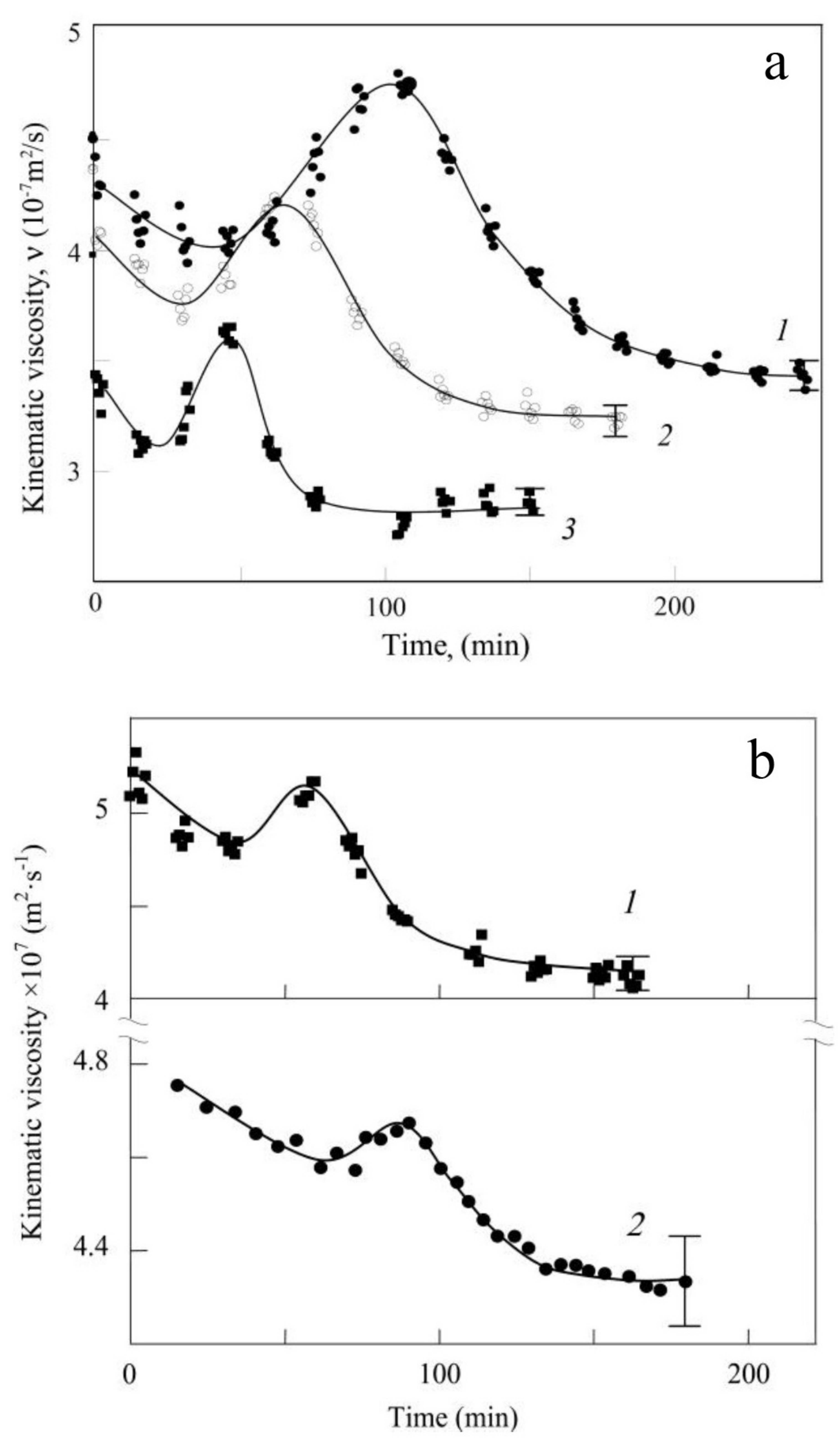}
   \caption{a --- The time dependencies of the liquid melt viscosity, Al$_{87}$Ni$_8$Y$_5$, at 900 $^{\circ}$C (1), 1050$^{\circ}$C (2) and 1200$^{\circ}$C (3), obtained after heating from room temperature \cite{Men}. b --- The time dependencies of the liquid melt viscosity, Al$_{86}$Ni$_8$La$_6$ (1) and Al$_{86}$Ni$_8$Ce$_6$ (2), at 1100$^{\circ}$C }
   \label{Experiment}
\end{figure}

\begin{figure}[h!]
   \centering
   \includegraphics[width=8cm]{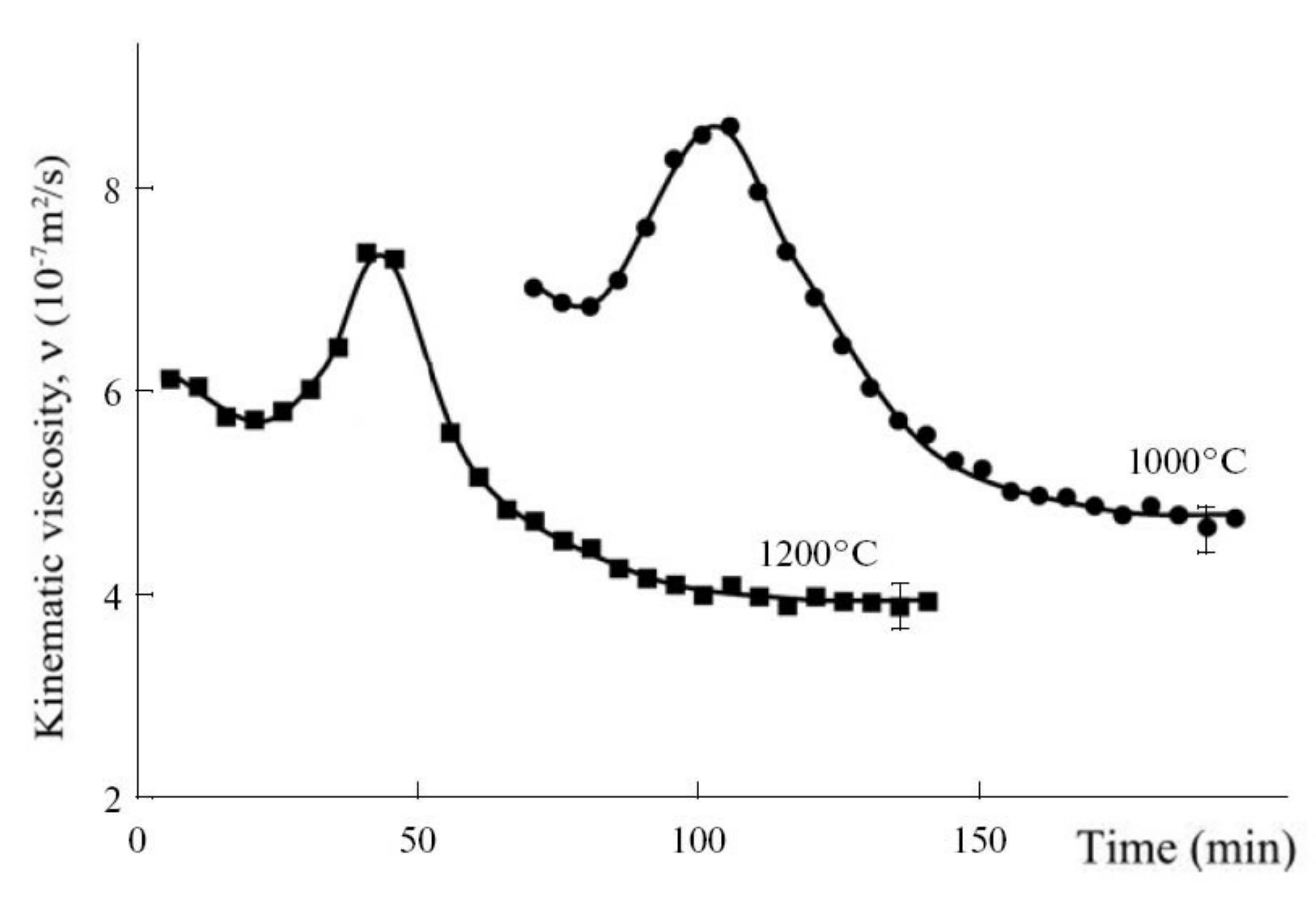}
   \caption{The temporal dependencies of the viscosity of Al$_{95}$Y$_5$ (a) and Al$_{90}$Y$_{10}$ (b) melts at the various temperatures \cite{Men,Belt}.}
   \label{A0}
\end{figure}

Moreover, in some cases these relaxation processes have unusual non-monotonic time dependence of viscosity \cite{Lad,Lad2,Men,Belt} (see Fig.\,\ref{Experiment} and Fig.\,\ref{A0}). Unusually is that in these experiments within some time interval after melting the melt viscosity demonstrates exponential decrease, but in some moment it unexpectedly starts to grow, reaches the local maximum, and then returns to the usual exponentially descending regime.

From the figures one can see, that the non-monotonic behavior of viscosity is observed in the melts of alloys of Al and Y with other impurities or without ones. Therefore, we can conclude that in general description of this phenomenon the basic point is the relaxation features of Al--Y melts, and, for simplification, focus to the description of the binary melt.

In our opinion the both phenomena (the long relaxation and the non-monotonic time dependence of viscosity) are related with the nonlinearity of the concentration dependence of the system chemical potential near to the solidus point. Weak nonlinearity of the system, i.e. the existence of two local minima of the free energy density, separated by small energy barrier, $E_{eb}<k_BT$, leads to the critically slow relaxation processes with fluctuation dynamics. These processes are caused by the complex cooperative motion of the system atoms. In homogeneous melts the amplitude of these fluctuations is small and cannot be observed experimentally. However, in case of strong initial inhomogeneity this inhomogeneity has sensitive influence on the experimental observed quantities like viscosity. It is very important that the lifetime of these inhomogeneities is determined by the characteristic correlation time of fluctuations. This correlation time can be calculated with help of the non-equilibrium critical dynamics methods \cite{Halp,Vas,Vasin} and can be macroscopically large. Below, using these methods, we will describe theoretically the slow non-monotonic relaxation processes discovered in Al--Y melts \cite{Belt}.

\section{Supposed physical nature of slow non-monotonic relaxation processes}
\label{sec:1}

We believe that the existence of extreme long relaxation time and non monotonicity of this relaxation in the discussed viscometric experiments is caused by the nonlinearity of the system, which arises as a result of the system finding in the critical regime near some critical point. We suppose that this point is the solidus point. Of course it is not the point of the second order phase transition, since the Y concentration is the conserved order parameter of the system, the total number of Y atoms in the system is limited, and any continuous phase transition is not possible in this situation. However, at the temperatures slightly above the solidus point the system has the pair of the free energy minima which correspond to low, and high Y concentrations. At low concentrations this minimum corresponds to the full Y dissolution in the liquid aluminum. At high concentrations the phase solution layering happens, and the Al--Y phase falls out in the melt, that corresponds to other minimum of the free energy. In the liquidus point these minima converge in one, that allows to believe that system dynamics near to this point is similar to the critical dynamics in vicinity of the critical point. But only this condition is not enough for the observation of the slow nonmonotonic relaxation.

The second condition is the inhomogeneous initial state of the system.
From the experiment we know that the initial alloy structure before the melting is inhomogeneous, since the inclusions of Al$_3$Y phase are present in the melt structure.
Usually the simple expressions describing the diffusion in liquid, are used for estimation of the dissociation time of these inhomogeneities. However, these expressions are correct only in the case, when the system's kinetics can be described within linear theory. The nonlinearity, caused by the features of the concentration dependence of the chemical potential, is able significantly change the diffusion kinetics.

%It would seem that the considered systems are at the temperatures significantly higher than the crystallization temperature, when the phase transition point is far. However, looking on the phase diagram of the Al--Y alloy, one can see that ...

\begin{figure}[h!]
   \centering
   \includegraphics[width=5cm]{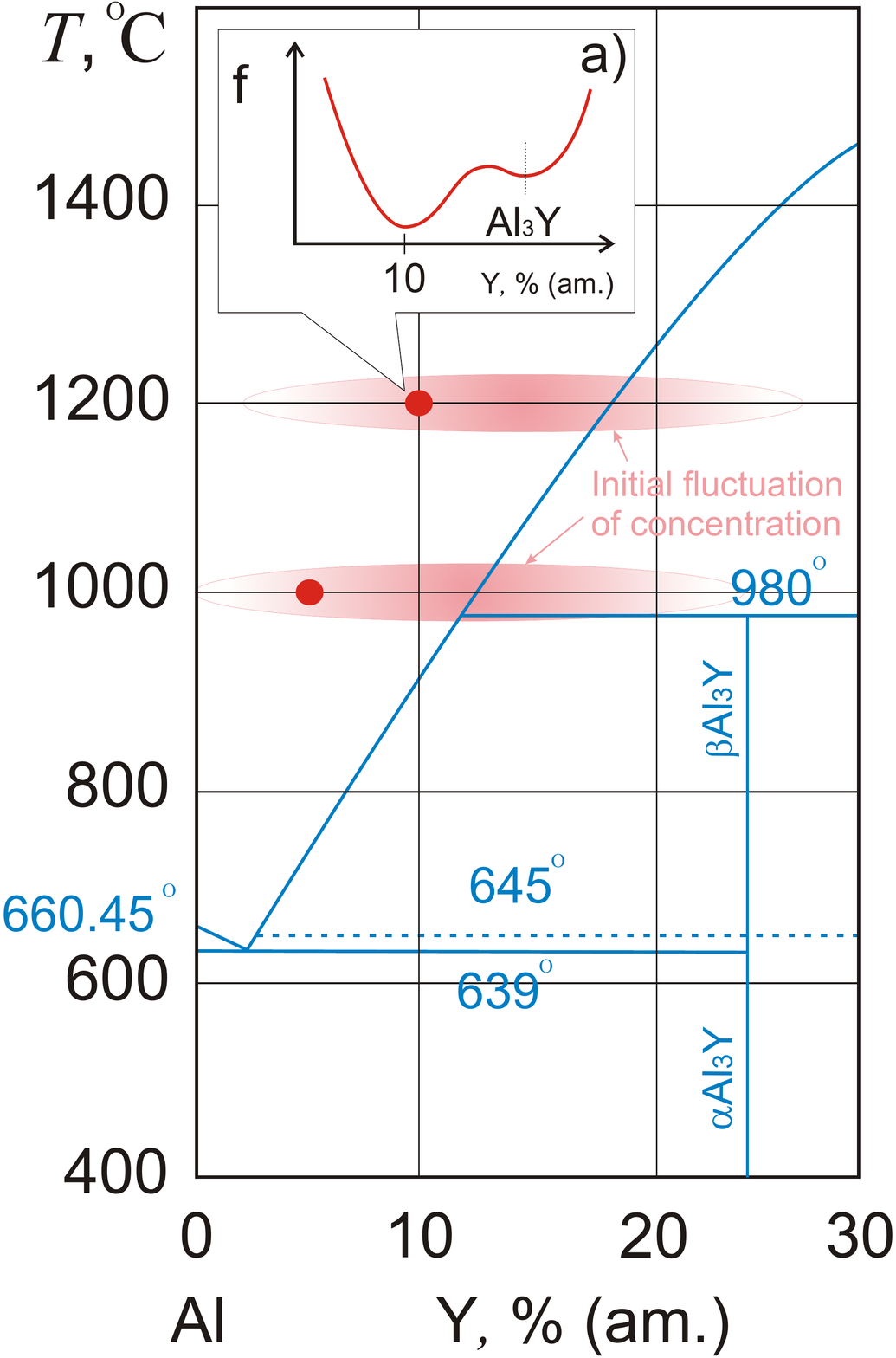}  %{PD2.pdf}
   \caption{We believe that the system is in the area of fluctuation behaviour, because the liquidus temperature line sharply increases with Y concentration, and even a slight deviation of Y concentration from the equilibrium value at the temperature belonging this interval, leads the system to the critical area of phase diagram. Sketch of the initial fluctuations area is pink. The red points correspond to the positions on the phase diagram of the equilibrium states of the melts considered in Fig.\,\ref{A0}. In a) we show the supposed qualitative form of the free energy density $f$ in which the metastable state with Al$_3$Y intermetallic complexes is present. }
   \label{FIG1}
\end{figure}

Thus we believe that the structural state of the system right after melting is spatially inhomogeneous in respect of the Y concentration.
Immediately after melting the destruction of these inclusions can not be described as a simple diffusion process, because the free energy of the system is non-linear on Y concentration. The Y atoms either go away from the areas with it's high concentration into the low concentration regions corresponding to the main free-energy minimum due to thermal fluctuations or come back to the high-concentration areas because of the local minimum of free energy density at high concentrations. Eventually the system will come to the equilibrium low-concentration state, but this process is nonlinear and the relaxation kinetics can be quite complicated.
Below we will try to understand, whether this nonlinearity can lead to the non-monotonic temporal dependence of viscosity, observed in the experiment.

\section{Model description}
For theoretical description of the observed effect we use the theory of phase separation.  We suppose that heterogeneous nucleation does not happen in the considered melt, and use the main provisions of the Cahn--Hilliard theory describing the kinetics of the homogeneous process of phase separation \cite{Cahn}.
In accordance with the Cahn--Hilliard theory the local concentration of Y in the melt, $\psi({\bf r})$, satisfies to the conservation law:
\begin{equation*}
\partial \psi({\bf r},\,t)/\partial t+\nabla {\bf j}({\bf r},\,t)=0,
\end{equation*}
where ${\bf j}$ is the concentration flow, which is proportional to the local difference between the chemical potentials $\mu ({\bf r})$, ${\bf j}=-\gamma \nabla \mu({\bf r})$, and $\gamma =D/T$ is the mobility. The chemical potential is depended by the free energy of the system:
\begin{equation*}
 \begin{array}{c}
\displaystyle F=\int d^3r\left[ f(\psi({\bf r}))+K(\nabla \psi({\bf r}))^2\right],\\[12pt]
\displaystyle \mu({\bf r})=\frac{\displaystyle \delta F}{\displaystyle \delta \psi({\bf r})}=\left( \frac{\displaystyle \partial f(\psi({\bf r}))}{\partial \psi({\bf r})}\right)_T-K\nabla^2\psi({\bf r}),
\end{array}
\end{equation*}
where $f$ is the free energy density, and the gradient term describes the contribution in system's energy, induced by the inhomogeneity. As a result, we come to the Cahn--Hilliard equation:
\begin{equation*}
\partial \psi({\bf r},\,t)/\partial t=\gamma  \nabla^2\left[ (\partial f(\psi)/\partial \psi)_T-K\nabla^2\psi\right].
\end{equation*}
This expression can be derived with help of the Keldysh technique, or other functional methods of non-equilibrium dynamics, if to consider the Y concentration as the order parameter of the system.
Above we suggest, that in some temperature interval above the solidus the system is in the critical fluctuations regime characterized by the long-range correlations.
The liquidus point is considered as the critical point, in which the pair of the free energy minima turns to the single minimum with the temperature decreasing.
For simplification we describe this system with help of  $\varphi^4$-model with conserved order parameter (H-model in \cite{Halp,Vas}) corresponding to the local Y concentration in our case.
Since we do not describe the heterogeneous nucleation in the melt, which is not possible at the considered Y concentration, then the terms with odd powers of $\varphi $ do not interest us, and we will not to take them into account.
Therefore, in our case $f({\bf r})=f_0+\dfr {m}2 \psi^2({\bf r})+\dfr{b}4 \psi^4({\bf r})$ ($\tau\leqslant 0$, $b>0$).
Below we consider the non-equilibrium dynamics of the system in fluctuation regime using the functional method.

\section{ Theoretical description  of the viscosity behaviour}
\label{DR} 
\begin{figure}[h!]
   \centering
   \includegraphics[width=7cm]{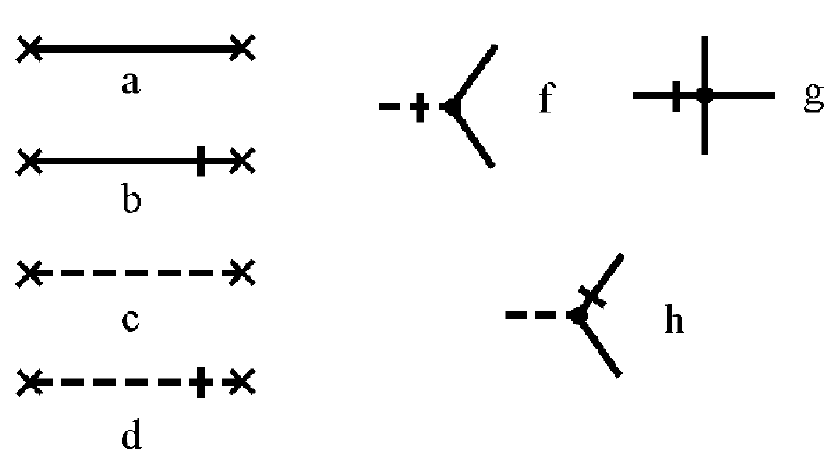}
   \caption{Graphs $a, b, c, d$  correspond to propagators $< \psi
\psi >$, $< \psi \psi '>$, $< v v >$, and $< v v'>$
 accordingly, graphs $f, g, h$  correspond to vertexes $v' \psi \partial (\partial ^2\psi )$, $\gamma b\psi '\partial
 ^2(\psi ^3)$, and $\psi 'v\partial \psi $  accordingly.}
   \label{fig4}
\end{figure}

Let us consider the fluctuations of the order parameter of metastable liquid in the region of phases co-existence. In order to describe the relaxation dynamics of the system close to the critical separation point we will consider the H-model~\cite{Halp,Vas}. This model is represented by the following equations system:
\begin{equation}\label{f5}
 \begin{array}{c}
   \displaystyle \dfr{\partial \psi}{\partial t}= \displaystyle \gamma   \nabla
^2\dfr{\delta F}{\delta \psi }-
   g_0\vec \nabla \psi
   \cdot \dfr{\delta F}{\delta \vec v}+\theta ,\\[12pt]
   \displaystyle \dfr{\partial \vec v}{\partial t}=
   \displaystyle P^{\perp}\left[ \; \eta _0 \nabla ^2 \dfr{\delta F}{\delta \vec
v}+
   g_0\vec \nabla \psi \dfr{\delta F}{\delta \psi }
   +\vec \xi \;\right] ,\\[12pt]
   F=\displaystyle \frac 12\int d^dx\left[ \;m \psi ^2+(\vec \nabla \psi
)^2+\frac {b}{2}\psi ^4+\vec
   v^2\;\right],
 \end{array}
\end{equation}
where $m =\alpha (T-T_{L})$, $T_L$ is the temperature of liquidus, $P^{\perp}$ is the projection operator which selects the transverse part of the
vector in brackets, and we accept that $K=1$. The first equation describes the dynamics of the order parameter $\psi $, and the second one describes the dynamics of the transverse part of the momentum density $v$, ($P^{\perp} $ is a projection operator which selects the transverse part of the vector in brackets, $\eta _0$ is the viscosity, $g_0=(\gamma \eta _0)^{-1}$ is the mode-coupling vertex, the functions $\vec \xi $ and $\theta $ are the Gaussian white noise source, which corresponds to thermal fluctuations of the system:
\begin{eqnarray*}
  \begin{array}{c}
   \langle \theta(x,\,t)\theta(x',\,t')\rangle=-\gamma  k_{B}T \nabla
   ^2\delta(x-x')\delta(t-t'),\\[12pt]
   \langle \xi _i(x,\,t)\xi _j(x',\,t')\rangle=-\eta _0 k_{B}T \nabla
   ^2\delta(x-x')\delta(t-t')\delta _{ij}.
  \end{array}
\end{eqnarray*}
These expressions ensure the fluctuation-dissipation theorem implementation. The critical properties of this model are known~\cite{Halp, Vas}. To analyze this stochastic model, one can employ the stochastic functional me\-thod \cite{Vas} and perturbation theory. Using these methods, one can write the field theory model corresponding our system, by a set of basic $\{ \psi ,\, v \} $ and supplementary $\{\psi ',\, v' \} $ fields, where the effective action will have the form of (Fig.\,\ref{fig4}):
\begin{equation*}
\begin{array}{l}
S (\Phi) = -\gamma  k_{B}T \psi ' \partial ^2 \psi ' + \psi ' [-
\partial _t \psi - \gamma   \partial ^2 ( \partial ^2 \psi - m \psi -b \psi
^3 )
-\\[12pt]
- v \partial \psi ]+ \gamma  ^{-1}k_{B}T g _{0} ^{-1} v ' \partial ^2
v ' + v ' [-\partial _t v + \gamma  ^{-1}  g _{0} ^{-1} \partial ^2
v + \psi
\partial (\partial ^2
\psi)].
\end{array}
\end{equation*}
The propagators of fields $\psi $ and $v$ have the form of:
\begin{equation*}
\begin{array}{c}
  \displaystyle \langle \psi \psi ^{\prime }\rangle =\langle \psi ^{\prime
} \psi \rangle
  ^T=\displaystyle \dfr 1{\varepsilon_k -i\omega
},\quad
  \langle \psi \psi \rangle =\dfr {2\gamma k_{B}T k^2}{ \left(
\varepsilon_k- i\omega
  \right) \left(\varepsilon_k +i \omega
\right)},\\[12pt]
  \displaystyle \langle v v^{\prime }\rangle =\langle v ^{\prime
}v\rangle
  ^T=\displaystyle \dfr {\gamma   g_{0}P^{\perp}}{
k^2-i\omega \gamma   g_0},\quad
  \langle v v\rangle =\dfr {2\gamma k_{B}T  g_{0}k^2}{ \left|
k^2-i\omega \gamma   g_0\right| ^2},
\end{array}
\end{equation*}
where
\begin{eqnarray*}
  \displaystyle \varepsilon _k=\gamma   k^2(k^2+M ),
\end{eqnarray*}
$M$ is the renormalized quantity $m$, and $k^2$ is the
impulse $\vec k$ squared.
Usually $G_R=\langle \psi \psi ^{\prime }\rangle$, $G_A=\langle \psi^{\prime } \psi\rangle$, and $G_K=\langle \psi \psi \rangle$ functions are called accordingly as retarded, advanced, and Keldysh Green function.

\begin{figure}[h!]
   \centering
   \includegraphics[width=3.7cm]{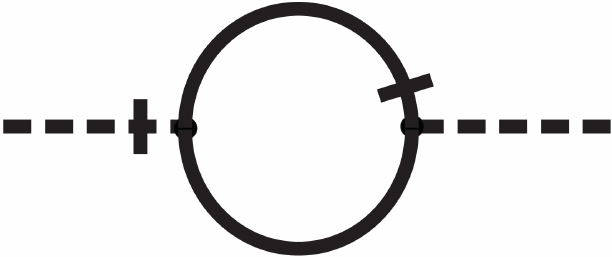}
   \caption{Graphical representation of the structural contribution into viscosity in one-loop approximation \cite{Vas}.}
   \label{Visc1}
\end{figure}

It is known \cite{Halp,Vas,Vasin} that the effective viscosity in this approach can be
represented as
\begin{eqnarray*}
  \eta ({\bf k},\,\omega )=\eta _0+\eta'({\bf k},\,\omega )=\displaystyle \eta _0\left[1 -\dfr {\displaystyle \gamma   g_0
  \Sigma _{v'v}({\bf k},\,\omega )}{\displaystyle k^2}\right] ,
\end{eqnarray*}
where $\Sigma _{v'v}$ is a coupled-mode contribution to the
response function, $\vec k$ is an external impulse, and $\eta'$ is the structural contribution into viscosity. In one-loop approach it can be represented in the diagram form (Fig.\,\ref{Visc1}), and has the analytical form
of:
\begin{eqnarray}
\label{V1}
\Sigma _{v'v}({\bf p},\,\omega )=\dfr {2k_{B}T}{d-1}\int \dfr{d{\bf k}}{(2\pi)^d}\dfr{k_iP^{\perp}_{ij}k_j\left( p^2-2\bf{pk}\right)}{\left( q^2+\tau \right)\left( \varepsilon_k+ \varepsilon_q -i\omega \right)}
\end{eqnarray}
where $\bf{q=p-k}$. In case of $d=4$ this integral contains the logarithmic divergence.
We believe this term dominates also in three-dimensional case.

Below we believe that the system relaxes very slowly in time. This corresponds to our experimental observations. In this case the time can be considered as a slow parameter of the system, which is constant in the renormalization procedure performed within the framework of the space scaling hypothesis.
From (\ref{V1}) one can see that the nonlinear additive to viscosity, in $({\bf r},\,t)$-representation, is the exponentially decaying function, $\Sigma _{v'v}(t)\propto e^{-t/t_{rel}}$, with relaxation time $t_{rel} \approx L^2M^{-1}\gamma  ^{-1} $, where $L$ is the characteristic initial correlation length. One can see that this relaxation time determines the melt viscosity value near the critical point.

\begin{figure}[h!]
   \centering
   \includegraphics[width=3cm]{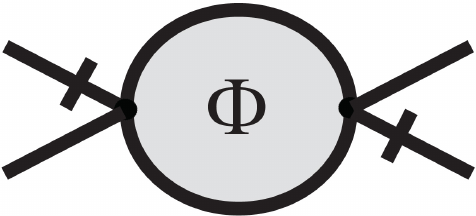}
   \caption{The diagrammatic representation of double-scattering process.}
   \label{A1}
\end{figure}

However, this approximation does not allow fully take into account the influence of complex cooperative processes on the relaxation kinetics, since in this approximation only integral effect of the fluctuation reduction processes is considered. In order to consider more detailed the time evolution of the melt viscosity one should also take into account the nonlinear processes of the double scattering of order parameter on the fluctuations. The diagrammatic representation one of the simplest of these processes is shown in fig.\,\ref{A1}. The double scattering corresponding to the reverse process, when the Y atom, left the cluster of the intermetallic complexes (we are considering them as the fluctuation), comes back to this cluster. These processes are not taken into account in the linear diffusion theory, but in critical regime they become important.
If the system is close to the critical point, then the diagrams with loops formed by the pair of the Keldysh parts of the Green function, like in fig.\,\ref{A1}, give the main contribution to the calculated values of the observed physical quantities.

\section{Non-monotonic time dependence of viscosity}

Below we consider the theoretical description of the non-monotonic time dependence of viscosity. We take into account the limitation of the size growth of the Al--Y complexes, believing that maximum value of the correlation length is $L$. Besides, we suppose that $M\neq 0$ and $L^{-1}<M^{1/2}$. Lastly we integrate of three-dimensional space.

Above we noted, that vicinity to the critical point the main contribution is given by the diagrams with the loop of the Keldysh correlators (see Fig.\,\ref{A1}).  One can show (see appendix I) that the loop, comprised in this diagram, contains the logarithmic divergence at $T=T_L$ in the case of $d=4$. In three-dimension case this diagram is significant too, and its analytical expression has the form of:
\begin{equation*}
\Phi({\bf k},\,t)\propto b^4\exp \left[-k^2(k^2+M)\gamma  |t|\right]\sqrt{\gamma  M|t|}
\end{equation*}
in $({\bf k},\,t)$-presentation (see Appendix I), and
\begin{equation*}
\Phi({\bf r},\,t)\propto \dfr{b^4}{M\gamma  |t|}\exp \left[-r^2/M\gamma  |t|\right]
\end{equation*}
in $({\bf r},\,t)$-presentation.

\begin{figure}[h!]
   \centering
   \includegraphics[width=7.5cm]{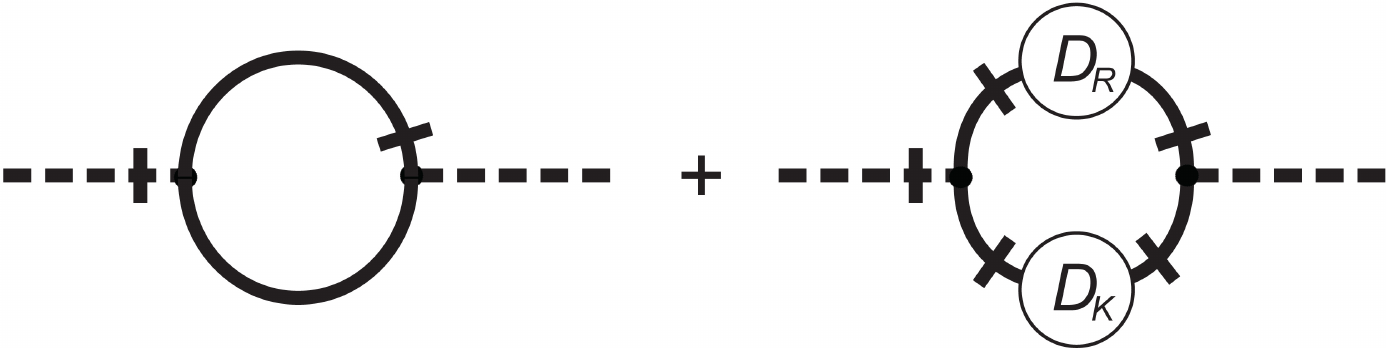}
   \caption{Structural contribution into viscosity, taking into account the double-scattering processes.}
   \label{R}
\end{figure}

However, this loop is not included in the one-loop approximation of the viscosity calculation (\ref{V1}). As it was noted above, this loop becomes significant when taking into account the double-scattering processes.
Therefore, we should to consider the contribution of this loop to the renormalization of the advanced, retarded, and Keldysh Green functions directly. Then the graphical representation of the viscosity expression has the form shown in fig.\,\ref{R}

\begin{figure}[h!]
   \centering
   \includegraphics[width=5cm]{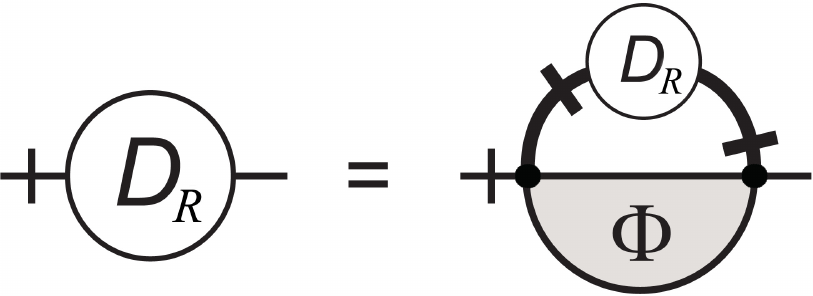}
   \caption{Graphical presentation of the self-consistent equation for $D_R$.}
   \label{A2}
\end{figure}

Let us consider the main contribution in the self-energy term, $D_R(k, \omega)$.
Taking into account that the loop gives dominant contribution to the renormalization of the vertex of this diagram, we neglect the contribution of the diagrams without these loops, and seek the self-consistent solution for the vertex as follows (Fig.\,\ref{A2}):
\begin{multline*}
\displaystyle D_R({\bf k},\,t)
=\theta(t)k^28\pi b^4L^4e^{-t/t_{rel}}\sqrt{2\pi |t|/t_{rel}}\\ \times L^{3}t_{rel}^{-1}\iint \theta(\tau)e^{-({\bf k}_1+{\bf k})^2(({\bf k}_1+{\bf k})^2+M)\gamma  \tau }D_R(t-\tau,\,{\bf k}+{\bf k}_1)k^2_1 d{\bf k}_1d\tau.
\end{multline*}
In assumption that $L^{-1}\ll M$, and in case of small $k$ ($k\to L^{-1}$),
\begin{multline*}
\displaystyle D_R(t)=\lim_{k\to L^{-1}}D_R(t,\,k)\propto \theta(t)\sqrt{t}\exp\left[4\pi b^4\sqrt{2\pi^2 }\,\mbox{erf}(\sqrt{t/t_{rel}})\right.\\
\left.-8\pi b^4\sqrt{2\pi |t|/t_{rel}}e^{-t/t_{rel}}-2t/t_{rel}\right]
\end{multline*}
(see Appendix II).

\begin{figure}[h!]
   \centering
   \includegraphics[width=5cm]{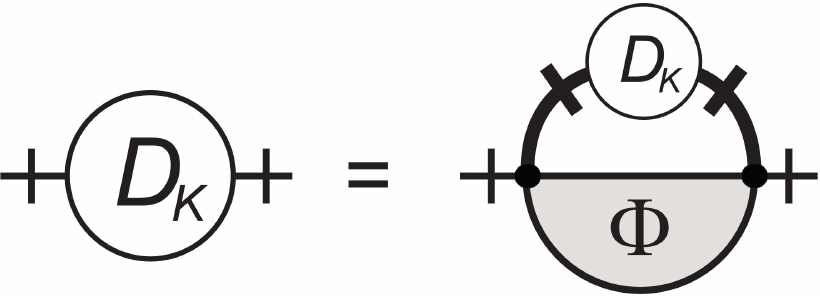}
   \caption{Graphical presentation of the self-consistent equation for $D_K$.}
   \label{A3}
\end{figure}

\begin{figure}[h!]
   \centering
   \includegraphics[width=9cm]{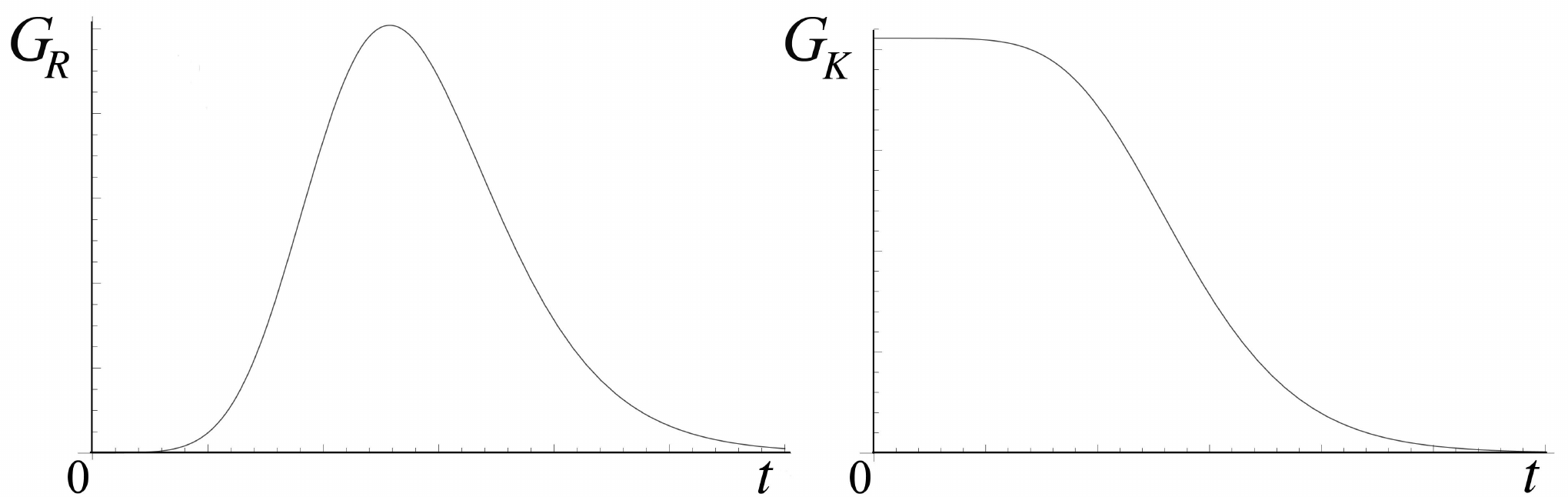}
   \caption{Qualitative form of the time dependence of $G_R$ and $G_K$ function.}
   \label{FF}
\end{figure}

\begin{figure}[h!]
   \centering
   \includegraphics[width=8cm]{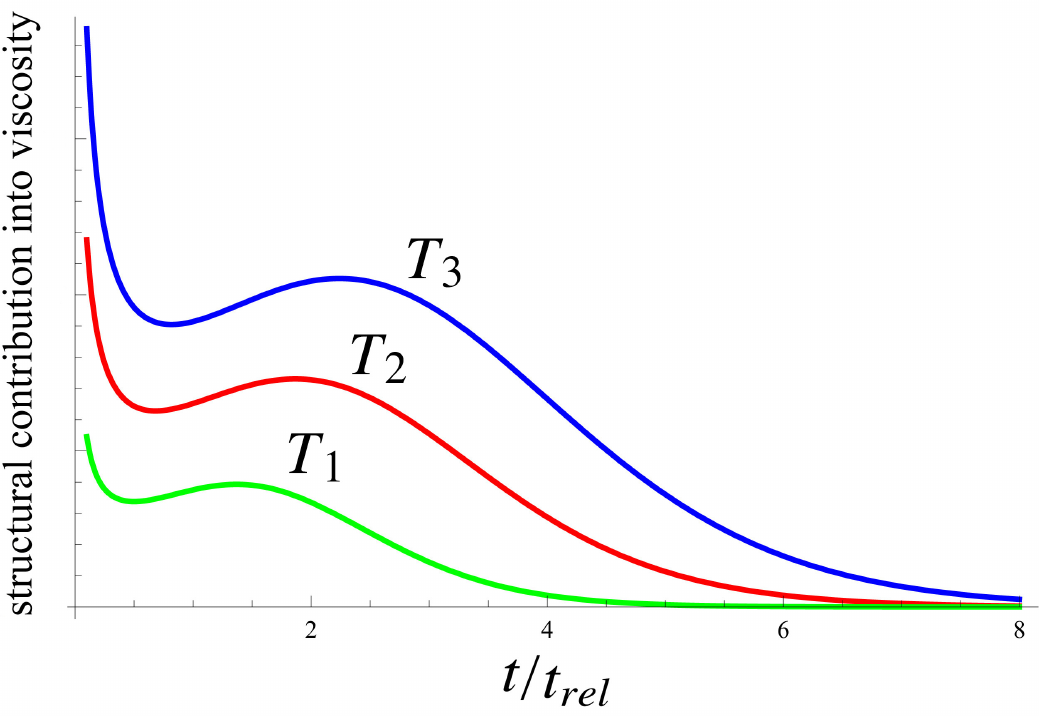}
   \caption{Qualitative form of calculated  time dependence of the structural part of the model viscosity at various temperatures ($T_1>T_2>T_3$).}
   \label{F1}
\end{figure}

Using this expression one can calculate the integral in the expression for the retarded correlation function in the loop:
\begin{multline*}
\displaystyle
G_R(t)=\int\limits_{-\infty}^{t} e^{-(t-\tau) /t_{rel}}D_R(\tau) d\tau \\ \propto\exp\left[4\pi b^4\sqrt{2\pi^2 }\,\mbox{erf}(\sqrt{t/t_{rel}})-8\pi b^4\sqrt{2\pi |t|/t_{rel}}e^{-t/t_{rel}}-t/t_{rel}\right].
\end{multline*}
Using the fluctuation-dissipation theorem,
\begin{equation*}
G_K(t)=\gamma k_{B}T \nabla^2\int\limits_t^{\infty}\left[ G_R(x)-G_A(x)\right] dx,
\end{equation*}
one can estimate the second factor:
\begin{multline*}
G_K(t)\approx\gamma k_{B}T L^{-2}\int\limits^{\infty}_{0} \int\limits_{t}^{\infty} e^{-y/t_{rel}}e^{-(x-t)/t_{rel}} D_R(x-y)dxdy \\ \propto\int\limits_{|t|}^{\infty}\exp\left[4\pi b^4\sqrt{2\pi^2 }\,\mbox{erf}(\sqrt{x/t_{rel}})\right.\\
\left. -8\pi b^4\sqrt{2\pi |x|/t_{rel}}e^{-x/t_{rel}}-x/t_{rel}\right] dx.
\end{multline*}
The qualitative graphical form of these functions is presented in Fig,\,\ref{FF}.
These expressions allow express the temporal dependence of the structural contribution into viscosity, $\eta'(t)$:
\begin{gather*}
\eta'(t)\approx \eta'_0\left[e^{-t/t_{rel}}+G_K(t)G_R(t)\right]
\end{gather*}
(Fig.\,\ref{F1}), where $\eta'_0$ is the time-independent factor.
The presented figures allow to see that the local maximum on the temporal dependence of viscosity is caused by the maximum on $D_{R}$ function, corresponding to the non-monotonic time dependence of susceptibility.

From the above analysis one can conclude that the nonmonotonicity of the relaxation processes is caused by the system's nonlinearity. The physical process of double-scattering, which generates this effect, is diagrammatically shown in Fig.\,\ref{A1}.  It can be interpreted as the return of Y atom in the high-concen\-tra\-tion region. This process is probable since it caused by the existence of the local minimum of free energy density. It is reflected in the local growth of the temporal function of response at $t<t_{rel}$ and as a result can lead to the temporal growth of the effective viscosity. Therefore, we believe that in the initial stage of relaxation this return process competes with the process of the diffusive dissociation of the initial concentration inhomogeneity and can lead to the observed nonmonotonicity.

\section{Conclusions}

According to the above arguments we can conclude, that the reason of the long-time relaxation and non-monotonic relaxation processes in two-component eutectic metal melts is the combination of two following factors:
\begin{enumerate}
  \item The concentration nonlinearity of the thermodynamic potential density near to the steep liquidus line above the eutectic, which leads to the critical slowing-down of the system's relaxation kinetics in the broad temperature interval. We believe that the temperature interval of critical behaviour is relatively wide, because the liquidus temperature sharply increases with Y concentration, and even a slight deviation of Y concentration from the equilibrium value at the temperature belonging this interval, leads the system to the critical area of phase diagram.
  \item Strong heterogeneity of the initial melt on concentration of Y atoms, which determines the initial amplitude of the concentration fluctuations, and which is sufficient to influence the measured properties of the melt.
\end{enumerate}

Our analysis allows to suggest the following scenario of the observed non-monotonic relaxation:
After melting of the initial solid alloy the resulting melt is strongly inhomogeneous. In this melt exist the regions with high concentration of the dissolved Y, and the regions with low concentration. The lifetime of these regions is relatively large, since the liquidus line is steep, and a small fluctuation of the local Y concentration can lead the system to the metastable state with high Y concentration.  These regions can contain non-dissolved intermetallic complexes Al--Y of various sizes.
Because of the Ostwald ripening in these regions small crystals or complexes dissolve, and redeposit into larger crystals or complexes~\cite{P1,P2,P4}.
Therefore, we suppose that in these high Y density areas the number of the segregated domains decreases, but their characteristic size grow in time. The increasing of the complexes sizes leads to the temporal increasing of viscosity.
At the same time, more slow process of these areas dissolution and homogenization of the melt also is passing.
As a result, the cluster growth process ends when Y concentration around of the complex becomes small, and the size of Al--Y phase critical nucleus, corresponding to this concentration, becomes large than the size of the intermetallic complex. After, when the Y concentrations in all regions level off and decreasing, the large complexes gradually dissolve. This leads to full homogenization of the melt and gradual decreasing of viscosity.

This work was supported by the Russian Foundation for Basic Research  (grants No. 14-02-00359 and No. 15-32-50291), M.G.Vasin is grateful to the Russian
Scientific Foundation (grant RNF 14-12-01185).

%\newpage
%$  $

\newpage

\section*{Appendix I}
We consider the system which is slow relaxing. The observation time of the system is close in value to the relaxation time, and we just are interested in determination of the temporal dependencies of physical values, i.e. time is considered as a parameter. Therefore, we will limit oneself, performing the renormalization procedure, by the consideration of only quasi-static approximation at the given time. In subsequent calculations we believe d=3. However, we take into account that main contributions in calculated values are given by the diagrams which are logarithmically divergent at d=4. In this case the main contribution in renormalization is given by the loops of two Keldysh Green functions, $\Phi $ (fig.\,\ref{A1}).

Since the form of the $\Phi(t,\,{\bf k})$ contribution is very important and has complicated form, it needs accuracy simplification in derivation.
The analytical form of this contribution is as follows:
\begin{equation*}
\begin{array}{l}
\displaystyle \Phi(t,\,{\bf k})= 2b^4L^3\int \dfr{e^{-(k_1^2(k_1^2+M)\gamma   +({\bf k}_1+{\bf k})^2(({\bf k}_1+{\bf k})^2+M)\gamma   )|t|}}{(k_1^2+M)(({\bf k}_1+{\bf k})^2+M)} d{\bf k}_1.
\end{array}
\end{equation*}

If we are interested in the description of the critical properties of the system ($M\to 0$), we should consider the epsilon expansion close to the critical dimensionality $d=4$.
Then the considered contribution logarithmically diverges at $k\to 0$ since $\Sigma \sim \int k^{-4}dk^4$. But we are interested in the description of the system, which is in the fluctuation regime, but not exactly in the critical point. We suppose that in this case the contribution of the loop also dominates at $k\to 0$, but the time dependence of this contribution should be calculated in three dimensional space.
In the case of $d=3$ for the small $k$, $k\ll M$, we have
\begin{multline*}
\displaystyle  \Phi(t,\,{\bf k})\approx -8\pi b^4L^3e^{-k^2(k^2+M)\gamma   |t|}\int\limits_{-\infty}^{\infty}\dfr{e^{-2k_1^2(k_1^2+M)\gamma   |t|}}
{(k_1^2+M)^2} k_1^2dk_1\\
\displaystyle \approx -8\pi b^4L^3e^{-k^2(k^2+M)\gamma   |t|}\int\limits_{-\infty}^{\infty}\left\{ \dfr{e^{-2k_1^2M\gamma   |t|}}
{k_1^2+M} -\dfr{Me^{-2k_1^2M\gamma   |t|}}
{(k_1^2+M)^2}\right\} dk_1\\
\displaystyle =-8\pi b^4L^3\left\{ \dfr{\pi(1+4\Gamma  M^2|t|)}
{2\sqrt{M}}e^{-(k^2+M)M\gamma   |t|}\,\mbox{erfc}\left[ \sqrt{2M\gamma   |t|}\right]\right.\\
\displaystyle \left. -e^{-(k^2+M)M\gamma   |t|}\,\sqrt{2\pi \gamma   M|t|}\right\}.
\end{multline*}
Therefore for $\Phi(t)=\lim_{k\to L^{-1}}\Phi(t,\,{\bf k})$ we have
\begin{multline*}
\Phi(t) \propto \lim_{k\to L^{-1}}\left(8\pi b^4L^3\exp \left[-k^2M\gamma   |t|\right]\sqrt{2\pi \gamma   M|t|}\right)\\
\approx 8\pi b^4L^4e^{-|t|/t_{rel}}\sqrt{2\pi |t|/t_{rel}},
\end{multline*}
where $t_{rel}=L^{2}M^{-1}\gamma  ^{-1} $ is the relaxation time.

\section*{Appendix II}
\label{sec:2}

Taking into account that the loop gives dominant contribution to the renormalization of the vertex of this diagram, we neglect the contribution of the diagrams without these loops, and seek the self-consistent solution for the vertex as follows (Fig.\,\ref{A2}):
\begin{multline*}
\displaystyle D_R(t,\,{\bf k})
=\theta(t)k^28\pi b^4L^4e^{-t/t_{rel}}\sqrt{2\pi |t|/t_{rel}}\\ \times L^{3}t_{rel}^{-1}\iint \theta(\tau)e^{-({\bf k}_1+{\bf k})^2(({\bf k}_1+{\bf k})^2+M)\Gamma  \tau }D_R(t-\tau,\,{\bf k}+{\bf k}_1)k^2_1 d{\bf k}_1d\tau\\
\approx \theta(t)k^28\pi b^4L^{2}e^{-t/t_{rel}}\sqrt{2\pi |t|/t_{rel}^3}\int\limits^{t}_{-\infty} e^{-k^2(k^2+M)\gamma   (t-x)} D_R(x,\,{\bf k})dx.
\end{multline*}
Differentiating this equation with respect to $t$,
\begin{multline*}
\displaystyle \partial_t D_R(t,\,{\bf k})\approx
8\pi b^4k^2L^{2}e^{-t/t_{rel}}\sqrt{2\pi |t|/t_{rel}^3}D_R(t,\,{\bf k})\\+8\pi b^4k^2L^{2}e^{-t/t_{rel}}\sqrt{2\pi |t|/t_{rel}^3}e^{-k^2(k^2+M)\gamma   t}\\
\times\left(\dfr{1}{2t}-k^2(k^2+M)\gamma  -1/t_{rel}\right)\int\limits^{t}_{-\infty} e^{k^2(k^2+M)\gamma   \tau } D_R(\tau,\,{\bf k})d\tau ,
\end{multline*}
we get the  following differential equation:
\begin{multline*}
\displaystyle \partial_t D_R(t,\,k)\approx \left(8\pi b^4k^2L^{2}\sqrt{2\pi |t|/t_{rel}^3}e^{-t/t_{rel}}\right.\\
\left. +\dfr{1}{2t}-k^2(k^2+M)\Gamma   -1/t_{rel}\right)D_R(t,\,k).
\end{multline*}
Up to a factor the solution of this differential equation is as follows:
\begin{multline*}
\displaystyle D_R(t,\,k)\approx C(k)\theta(t)\sqrt{|t|}\exp\left[4k^2\pi b^4L^{2}\sqrt{2\pi^2}\,\mbox{erf}(\sqrt{t/t_{rel}})\right.\\
\left.-8k^2\pi b^4L^{2}\sqrt{2\pi |t|/t_{rel}}e^{-t/t_{rel}}-(k^2(k^2+M)\gamma   +1/t_{rel})t\right].
\end{multline*}
where $C(k)$ does not depend on $t$. At small $k$ ($k\to L^{-1}$)
\begin{multline*}
\displaystyle D_R(t)=\lim_{k\to L^{-1}}D_R(t,\,k)\propto \theta(t)\sqrt{|t|}\exp\left[4\pi b^4\sqrt{2\pi^2 }\,\mbox{erf}(\sqrt{t/t_{rel}})\right.\\
\left.-8\pi b^4\sqrt{2\pi |t|/t_{rel}}e^{-t/t_{rel}}-2t/t_{rel}\right].
\end{multline*}

Using this expression one can calculate the integral in the expression for the correlation function:
\begin{multline*}
\displaystyle
\int \theta(\tau)e^{-k^2(k^2+M)\gamma   \tau }D_R(t-\tau,\,{\bf k}) d\tau \\
=\dfr {t_{rel}e^{t/t_{rel}}}{8\pi b^4k^2 L^{2}\sqrt{2\pi |t|/t_{rel}}}D_R(t,\,{\bf k})\\
\propto\dfr { \theta(t)\sqrt{t_{rel}^3}}{8\pi b^4\sqrt{2\pi}}\exp\left[4\pi b^4\sqrt{2\pi^2 }\,\mbox{erf}(\sqrt{t/t_{rel}})\right.\\
\left.-8\pi b^4\sqrt{2\pi |t|/t_{rel}}e^{-t/t_{rel}}-t/t_{rel}\right].
\end{multline*}
Therefore
\begin{multline*}
\displaystyle
\int\limits_{-\infty}^{t} e^{-(t-\tau) /t_{rel}}D_R(\tau) d\tau \\
\propto\exp\left[4\pi b^4\sqrt{2\pi^2 }\,\mbox{erf}(\sqrt{t/t_{rel}})-8\pi b^4\sqrt{2\pi |t|/t_{rel}}e^{-t/t_{rel}}-t/t_{rel}\right].
\end{multline*}

\end{document}